\documentclass[twocolumn,prb]{revtex4}
\usepackage{amsmath}
\usepackage{amsfonts}
\usepackage{amssymb}
\usepackage{graphics}
\usepackage{float}
\usepackage{graphicx}
\usepackage{epstopdf}
\usepackage{pifont}
\usepackage[compact]{titlesec}
\usepackage{natbib}
\usepackage{threeparttable}
\usepackage{lipsum}

\begin{document}
\title{Atomistic and continuum modeling of a zincblende quantum dot heterostructure}
\author{Parijat Sengupta}
\email{psengupta@purdue.edu}
\author{Sunhee Lee}
\author{Sebastian Steiger}
\author{Hoon Ryu}
\author{Gerhard Klimeck}
\affiliation{Dept of Electrical and Computer Engineering, Purdue University, West
Lafayette, IN, 47907 
}

\begin{abstract}
A multiscale approach was adopted for the calculation of confined states in self-assembled semiconductor quantum dots (QDs). While results close to experimental data have been obtained with a combination of atomistic strain and tight-binding (TB) electronic structure description for the confined quantum states in the QD, the TB calculation requires substantial computational resources. To alleviate this problem an integrated approach was adopted to compute the energy states from a continuum 8-band k.p Hamiltonian under the influence of an atomistic strain field. Such multi-scale simulations yield a roughly six-fold faster simulation. Atomic-resolution strain is added to the k.p Hamiltonian through interpolation onto a coarser continuum grid. Sufficient numerical accuracy is obtained by the multi-scale approach. Optical transition wavelengths are within 7$\%$ of the corresponding TB results with a proper splitting of p-type sub-bands. The systematically lower emission wavelengths in k.p are attributable to an underestimation of the coupling between the conduction and valence bands.

\end{abstract}
\maketitle

\section{Introduction}
\label{intro} 
Self-assembled semiconductor quantum dots(SAQDs) are being employed as active meta-materials in high-speed semiconductor lasers, which achieve high-speed data transmission while utilizing minimal power. SAQDs are also one of the simplest means of exploring the physics of carriers in a three-dimensional confined regime. In this contribution the electronic properties of such dots are explored through atomistic, continuum and a combination of both approaches. Traditionally quantum dots have been modelled within an effective mass approach using the multi-band k.p formulation~\cite{grundmann1995inas}. While the results obtained through these methods work well, the details of the atomic arrangement are disregarded~\cite{wang1999electronic}. 

To preserve the microscopic character of the dot and its interfaces, the atomistic tight-binding approach can be adopted for energy calculations. Local strain, which considerably modifies the energy spectrum, is accounted for through the atomistic valence force field (VFF) method~\cite{musgrave1962general}. Results from combinations of these methods are presented and analysed here regarding their symmetry and their deviation to experiment. After a brief summary of the methods and a description of the studied QD, some numerical aspects are highlighted. The shortcomings of some of these methods are demonstrated and it is shown that their origin lies in the assertion upon which the theory is constructed.

\section{Model and Method}
\label{meta}

In this section the strain and electronic structure models are briefly reviewed, combinations of which are applied to the structure described in the following section.
The atomistic TB Hamiltonian is constructed by choosing a relevant set of orbitals localized on an atom. The wave function of the system is expanded in the basis of these localized orbitals. The inter- and intra-atomic matrix elements which enter the Hamiltonian are empirically fitted to the dispersion, band gaps and masses~\cite{slater1954simplified,harrison2012electronic}. The obtained results exhibit by construction the same symmetries as the underlying crystal. The total Hamiltonian of a nanostructure is constructed from its constituent bulk material parameters.
Continuum approaches using multi-band k.p theory rest on the assumption of a slowly varying envelope function and reduce to the simple effective mass model in the case of a single band. The eight-band model~\cite{luttinger1955motion,kane1957band,bahder1990eight,liu2002modeling} for zincblende is used for continuum calculations of QD eigenstates. It includes two s-type conduction band and six valence bands. Strain is added through the standard Bir-Pikus deformation potentials.

The atomistic valence force field (VFF) model for lattice properties in its simplest form due to Keating~\cite{keating1966effect} expresses the energy of the crystal lattice through a two-body term that describes bond stretching and a three-body interaction that reports the bond bending (angular distortion). The total crystal energy is obtained by summing over all the atoms in the domain and their corresponding nearest neighbours. This work augments the Keating force constants by an anharmonic dependence which has been shown to be necessary in the case of non-vanishing strain~\cite{usman2009moving}. This observation will be again made in the present work.

\begin{figure}
\includegraphics[scale= 0.8]{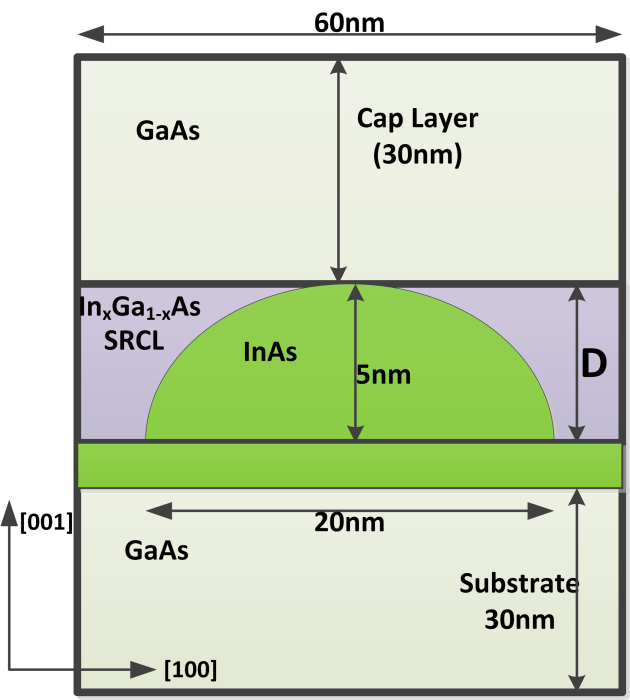}
\caption{Cross section through the modeled self-assembled quantum dot. The InAs dot itself is dome-shaped and embedded in an InGaAs stress-reducing layer (SRCL) with mole fraction x=0.4. The structure is embedded in a GaAs matrix.} 
\label{fig1}
\end{figure}

Figure ~\ref{fig1} shows the schematic of the simulated system. 10 A dome-shaped InAs QD of 5nm height and 20 nm base diameter sits on a single monolayer of InAs that serves as the wetting layer. The QD is set in an InxGa1-xAs alloy of mole fraction 40$\%$ which functions as the stress-reducing layer (SRCL)~\cite{tatebayashi2001over}. The height of the SRCL is 5 $ \mathrm{nm}$. A GaAs host matrix (60 $\mathrm{nm}$ x 60 $ \mathrm {nm}$ x 60 $ \mathrm{nm}$) surrounds the whole structure. An eight-band k.p model with parameters~\cite{chuang2012physics} fitted to bulk band structure was employed for continuum calculations while atomistic calculations are performed by utilizing a 20-band sp$^{3}$d$^{5}$s$^{*}$ TB model~\cite{boykin2004valence}. Spurious solutions in the 8-band k.p model are removed by adjusting Kane’s parameter~\cite{veprek2007ellipticity}. Of the several inter-atomic potentials~\cite{stillinger1985computer,tersoff1989modeling} that can be used to set-up a VFF calculation, the popular Keating model is used, though refinements have been worked out~\cite{bernard1991strain}.  

\section{Numerical Aspects}
In this section some computational aspects particular to this work are highlighted. The reader is referred elsewhere for a more complete discussion~\cite{lee2009million}. The presented results were obtained with an extended version of NEMO-3D-Peta~\cite{ryu2012full}. To interpolate the atomistic strain field obtained through a VFF approach (Table.~\ref{table1}) onto the continuum grid used for the k.p calculation, a sphere was constructed around each continuum node and the average over all atoms within this sphere was taken for the strain tensor. To reduce compute times by utilization of hundreds to thousands of computing cores in parallel, a 3D-domain decomposition scheme is employed. A device can be spatially decomposed into three dimensions and each sub-domain is assigned to a corresponding CPU. Based on the spatial information, each CPU only stores the information of the atoms in its sub-domain and neighbouring atoms from adjacent sub-domains; no global position information is held locally, minimizing memory consumption and enabling the simulation of large devices.
\begin{table}[!htb]
\centering
\caption{Scheme for interpolated VFF based calculations. Continuum strain cannot be combined with TB calculations and is indicated by a \ding{55}  in the table.}
\label{table1}
\begin{tabular}{*3c}
\toprule
Electronic Structure & \multicolumn{2}{c}{Strain Model} \\ \hline
 & Keating VFF & Continuum Strain \\
sp$^{3}$d$^{5}$s$^{*}$  TB & \ding{51} & \ding{55} \\
8-band k.p  &  \ding{51} & \ding{51} \\ \hline
\end{tabular}
\end{table}
 
The major drawback for 3D- parallelization is the increase of the complexity of communication, which may cause significant performance degradation; there is a trade-off between reducing the computational burden and increasing communication overhead. 
Fifteen million atomic positions were considered in the computation of the VFF-based strain components. TB simulation is performed on a subset of nine million atomic positions. The continuum model calculations used a homogeneous grid with 1 nm spacing that resulted in a total of 216,000 nodes. As a result calculations are roughly six times faster in the continuum case over their atomistic counterpart. An even larger speed-up is prevented by a loss of sparsity in the continuum model Hamiltonians.

\section{Results}

\begin{figure}
\includegraphics[scale= 0.75]{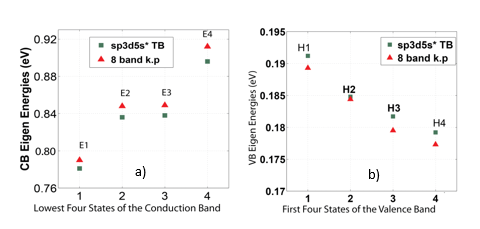}
\caption{Comparison between atomistic tight-binding and continuum k.p results obtained for the first four confined a) electron and b) hole energy levels. These calculations are done without the influence of strain.} 
\label{fig2}
\end{figure}

In this section the performance and accuracy of the atomistic and continuum methods applied to the QD sketched in Fig.~\ref{fig1} is presented. In Fig.~\ref{fig2} a comparison of the first conduction and valence band states is shown in the absence of strain.
While the electron and hole energy states show reasonable quantitative agreement, there exists non-degeneracy for the excited states in the atomistic case. The underlying atomic structure of the zincblende lattice has, on account of loss of inversion symmetry, a reduced symmetry (C$_{2v}$ point group) but the continuum model sees a higher C$_{4v}$ symmetry. The $ n=2,3 $ k.p states are energetically degenerate and linear combinations of p-like states. Tight binding resolves the quantum dot's microscopic foundation and any additional structural asymmetry attributable to interface roughness and asymmetry thus breaching the artificial degeneracy of the continuum model. Atomistic strain effects which include bond stretching and rotation at each atomic position are naturally incorporated in a TB description. Therefore the C$_{4v}$ symmetry that preserves the equivalence along [110] is lost in the zincblende crystal and consequently so is the degeneracy. 

\begin{figure*}
\includegraphics[scale= 1]{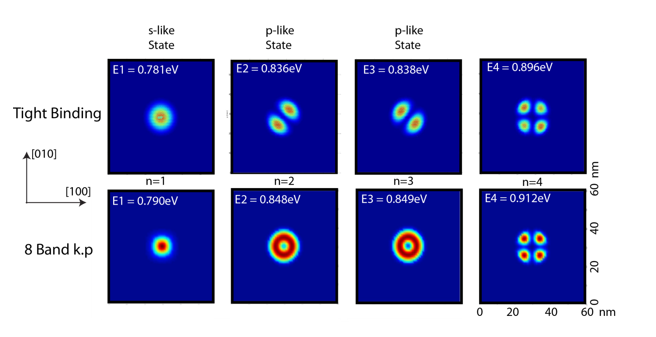}
\caption{Anisotropy of wave functions and non-degeneracy of states for $ n = 2, 3 $ for a tight binding calculation. Corresponding k.p wave-functions are radially symmetric.} 
\label{fig3}
\end{figure*}

This also sets up an inaccurate portrayal of symmetry-dependent effects such as anisotropy of wave functions and level anti-crossing in addition to loss of degeneracy. In Fig.~\ref{fig3} the wave functions and the corresponding energies of the conduction band states seen in Fig.~\ref{fig2} are plotted. Due to the non-splitting of the p-level sub-bands in k.p, continuum methods fail to reproduce the polarization anisotropy in electron-hole transition matrix elements.

While theoretically the $n = 2,3 $ states should produce degenerate states under a k.p calculation, we note from Fig.~\ref{fig3} that a splitting of 1 $ \mathrm{meV}$ remains in our continuum calculations. This is attributed to numerical errors including the coarseness of the finite difference mesh.
The results above were computed in the absence of strain and show considerable divergence from experimentally gathered data, as can be seen in Table ~\ref{table2}. Strain fields, which modify the band edge and are significant within and in the vicinity of the dot, were computed using the VFF method and added to the TB Hamiltonian~\cite{luisier2006atomistic}.

\begin{figure}
\includegraphics[scale= 0.6]{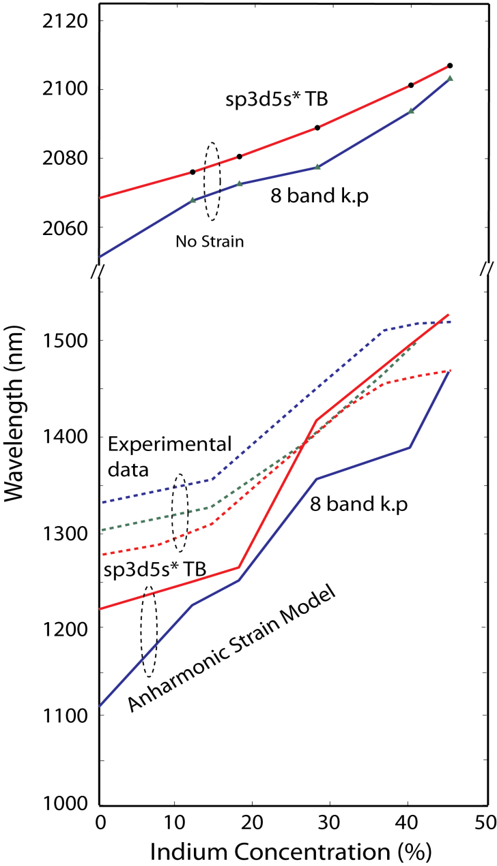}
\caption{Optical Transition wavelength vs. Indium conc. A much closer match is obtained with experimental data for an anharmonic VFF model combined with the 8-band continuum k.p.} 
\label{fig4}
\end{figure}

From Table.~\ref{table2}, it is seen that a while a simulation under a strain model with a harmonic approximation gets the results closer to the laboratory data, the neglect of anharmonic modifications of the inter-atomic potential leaves ample room for improvement. Anharmonicity is added to the inter-atomic potential by utilizing distance- and angle-dependent VFF constants~\cite{lazarenkova2004effect}. By inclusion of anharmonic corrections in the simulation, the results are within 4-9 $\%$ of reported experimental values (see Table~\ref{table2} and Fig.~\ref{fig4}). Another fact supporting the inclusion of anharmonicity, a non-linear dependence of the emission wavelengths as a function of the SRCL mole fraction which is in agreement with experiment is not discussed here.Non-linearity inherent in the atomistic description is not captured well by continuum k.p as borne out by Fig.~\ref{fig4}. Further corrections due to piezoelectric effects operational in polar GaAs were not considered in this work.
\section{Conclusions} 
A comparison of continuum and atomistic methods for the calculation of energy states in a realistically extended quantum dot is presented. The multi-band k.p model with a limited basis set cannot exactly predict results arising out of the atomistic nature of interfaces and lowered symmetry of underlying crystal until the internal symmetry is incorporated in to it through an atomistic model. This is achieved by introducing a symmetry-breaking strain calculation using the valence force field model. The multi-scale combination of atomistic strain and continuum band structure offers six-fold faster and reasonably accurate results to compute the electronic structure of large QDs. Specifically, inclusion of VFF strain recovers the p-state anisotropy in k.p calculations and yields emission energies fairly close to atomistic predictions.

\begin{table}
\caption{Transition energies between the first confined electron and hole states for various models, and experimental values.}
\centering
\label{table2}
\begin{tabular}{lcc}
\noalign{\smallskip} \hline \hline \noalign{\smallskip}
Model & E1-H1 transition energy ($\mathrm{eV}$) \\\hline
Experiment~\cite{tatebayashi2001over} & 0.81 - 0.85\\
k.p unstrained  & 0.594 \\
TB unstrained & 0.591 \\
k.p strained (harmonic VFF)  & 1.063 \\
TB strained (harmonic VFF)  & 1.040 \\
k.p strained (anharmonic VFF) & 0.885 \\
TB strained (anharmonic VFF) & 0.828 \\
\noalign{\smallskip} \hline \noalign{\smallskip}
\end{tabular}
\end{table}

\begin{acknowledgements}
The authors would like to thank Muhammad Usman for useful discussions. This work was financially supported by NSF grant 102423. Computational resources of nanoHUB.org, hosted by the Network for Computational Nanotechnology (NCN), are acknowledged.
\end{acknowledgements}

\bibliographystyle{apsrev}
\bibliography{References}

\end{document}